\documentclass[11pt, letterpaper]{article}
\usepackage{amsfonts}
\usepackage{amsmath}
\usepackage{amssymb}
\usepackage{enumerate}
\usepackage{natbib}
\usepackage{color}
\usepackage{array}
\usepackage{graphicx,epstopdf}
 \usepackage{float}
\usepackage{rotating}
\usepackage{url}
\usepackage{subfigure}
\usepackage{caption}
\usepackage{setspace}
\usepackage{hyperref}
\newtheorem{theorem}{Theorem}

\newtheorem{remark}[theorem]{Remark}

\def\R{{\mathbb R}}        
\def\1{{\mathbf 1}}        

\begin{document}
\title{The Golden Target:\\ Analyzing  the Tracking Performance of Leveraged 
Gold   ETFs }
\author{Tim Leung\thanks{Industrial Engineering \& Operations Research (IEOR) Department, Columbia University, New York, NY 10027,  email: {tl2497@columbia.edu}. Corresponding author. } \and Brian Ward\thanks{Industrial Engineering \& Operations Research (IEOR) Department, Columbia University, New York, NY 10027, email: {bmw2150@columbia.edu}.} }

\maketitle

\abstract{This paper studies  the  empirical tracking performance of  leveraged  ETFs on gold, and    their price  relationships  with gold spot and futures.  For tracking the gold spot, we find that our optimized  portfolios with short-term gold futures      are  highly effective in replicating prices. The market-traded gold ETF (GLD) also exhibits a similar tracking performance. However, we show  that   leveraged   gold ETFs tend to underperform their corresponding  leveraged benchmark. Moreover,  the underperformance  worsens  over a longer holding period. In contrast, we illustrate that a dynamic portfolio of gold futures tracks  significantly better than various static portfolios. The dynamic portfolio also consistently outperforms the respective  market-traded LETFs  for different leverage ratios over  multiple years.    }

\tableofcontents
\newpage 
\section{Introduction}\label{Introduction}
Gold is often viewed by investors as a safe haven or a hedge against market turmoils, currency depreciation, and other  economic or political events.\footnote{See \cite{goldhedge}  and 
\cite{BaurThomas} for empirical investigation of gold's role as a safe haven.} For instance, during 2008-2009, major market indices, including the Dow and the S\&P 500,   declined by about  20\% while gold prices rose from \$850 to \$1,100 per troy ounce. In August 2011, the price of gold reached a peak of \$1,900 an ounce soon after the Standard \& Poor's downgrade of the U.S. credit rating. However, a direct investment in gold bullion is difficult for many investors due to high storage cost and the lack of a liquid exchange for gold bullions. In order to gain exposure to gold, investors can alternatively trade various instruments, such as gold futures,   gold exchange-traded funds (ETFs), exchange-traded notes (ETNs), and  leveraged ETFs/ETNs. 

 Gold futures are exchange-traded contracts written on 100 troy ounces of gold, with a number of available delivery dates within 5 years of any given trading date. In the US, gold futures are traded at the New York Mercantile Exchange (NYMEX). The available months include the front three months, every February, April, August and October falling within the next 23 months, and every June and December falling within the next 72 months. Trading for any specific contract terminates on the third to last business day of the delivery month.\footnote{Historical quotes and contract specifics of gold futures are obtained from the CME Group (http://www.cmegroup.com/trading/metals/precious/gold.html).}  As we will see, the futures prices are highly correlated among each other and the various gold (L)ETFs. 

 Gold ETFs and ETNs are designed to track the spot price of gold, and are liquidly traded on exchanges like stocks. In fact, the SPDR Gold Trust ETF (GLD), is one of the most traded ETFs with an average trading volume of 6.2 million shares and market capitalization of US \$33 billion as of July 2014.\footnote{According to ETF Database (http://www.etfdb.com/compare/volume).} Within this gold ETF market, there are funds which seek to provide investors with a return equal to a constant multiple of the daily returns of spot gold. Such funds are called leveraged ETFs (LETFs). Common leverage ratios are $\pm2$ and $\pm 3$ and LETFs usually charge an expense fee for the service. Major issuers include ProShares, iShares, and VelocityShares (see Table \ref{tab:Data Table}). For example, the VelocityShares 3x Long Gold ETN (UGLD) provides a return of 3 times the gold spot price. Furthermore, one can take a bearish position on the gold spot price by investing in an LETF with a negative leverage ratio. An example  is the VelocityShares 3x Inverse Gold ETN (DGLD).

 In this paper, we investigate the price dynamics of the various financial products related to gold described above. In Section \ref{futures}, we examine through a series of regressions the price relationships between the futures contracts and spot gold.  We find significant price co-movements among them, and their returns tend to be closer over a longer holding period. Furthermore, we construct static portfolios consisting 1 or 2 futures with different maturities to replicate the spot gold price. By comparing the tracking performance of these portfolios, we find that the 1-month futures is most useful for replicating the gold spot price. 
 
 In Section \ref{LETFs}, we discuss the price dynamics of  leveraged gold  ETFs, and their tracking  performance against their leveraged benchmark. For all gold LETFs studied herein, their average returns tend to be lower than the corresponding multiple of the  reference's returns, and the underperformance worsens  as the holding period increases.  For tracking the leveraged benchmark, we find that static replication with futures is ineffective. Therefore, we construct a dynamic leveraged  portfolio using the 1-month futures.  Over a long  out-of-sample period, we demonstrate that  this portfolio tracks the leveraged benchmark better than the corresponding LETFs.  

\begin{table}[th] \centering\begin{small}
    \setlength{\extrarowheight}{2pt}
    \begin{tabular}{l l l l  r l l }  
        \hline
        \text{LETF} & \text{Reference} & \text{Underlying} & \text{Issuer}   & \text{$\beta$} & \text{Fee} & \text{Inception} \\ \hline
        \hline
        GLD                      & GOLDLNPM                            & Gold Bullion         & iShares       	& 1\,       & 0.40\%   & 11/18/2004 \\
        UGL                      & GOLDLNPM                            & Gold Bullion         & ProShares     	& 2\,       & 0.95\%   & 12/01/2008 \\
        GLL                      & GOLDLNPM                            & Gold Bullion          & ProShares     	& $-$2\,      & 0.95\%   & 12/01/2008 \\
        UGLD                   & SPGSGCP                             & Gold Bullion           & VelocityShares	& 3\,       & 1.35\%   & 10/17/2011  \\
        DGLD                   & SPGSGCP                             & Gold Bullion           & VelocityShares	& $-$3\,      & 1.35\%   & 10/17/2011 \\
       \hline    \end{tabular}  \end{small}
     \parbox{0.8\textwidth}{\caption{\small{A summary of the  gold LETFs, along with the unleveraged ETF (GLD). The LETFs with higher absolute leverage ratios, $|\beta|\in\{2, 3\}$,  tend to have higher expense fees. }}\label{tab:Data Table}} 
\end{table}


Among related studies on gold ETFs, \cite{baurFreeLunch} examines the cost-effectiveness of gold ETFs relative to physical gold holdings, and discuss the effect of commodity financialization on the prices of gold and associated ETFs. \cite{ivanov} uses t-tests to study the tracking errors for gold, silver, and oil (unleveraged) ETFs. Using prices from March-August 2009, he concludes that such ETFs closely track their underlying assets. In contrast, we consider the tracking performance of  leveraged gold ETFs over multiple years, and construct dynamic futures portfolio to replicate the leveraged benchmark based on spot gold. 

Futures are an important instrument for hedging commodities (L)ETFs as these funds are often fully or partially  constructed using  futures and swaps contracts rather than the physical asset. \cite{futures} provides an overview of commodity ETFs constructed with futures. \cite{AB} discuss hedging strategies with futures contracts for index ETFs and compare them against some minimum variance hedge ratios. Empirical studies  by \cite{smales} and \cite{baurGoldVol} show  that the volatility of gold spot and futures exhibits asymmetric responses to market shocks. The study argues that the higher sensitivity of gold volatility to positive shocks can be interpreted by the safe haven property of gold.

%

There are a number of  studies on the price dynamics of   LETFs in general, including   \cite{cheng},  \cite{AZ}, and \cite{jarrowLETFs}.  They illustrate how  the return of an LETF can erode proportional to  the leverage ratio as well as the realized variance of the reference index.  For equity ETFs,  \cite{RompotisSEF} applies regression to determine the  tracking errors between ETFs and their stated benchmarks, and finds persistence in  tracking errors over time. The horizon effect is also   illustrated in the empirical study by \cite{Murphy}  for  commodity  LETFs. \cite{guoleung} systematically study the tracking errors of a large collection of commodity LETFs. They define a realized effective fee to capture how much an LETF holder  effectively  pays the  issuer due to the fund's tracking errors. Furthermore, \cite{HLMM} corroborate the volatility effect by using VIX data in a linear regression of the returns. They also find that the change in the expected volatility is even more significant in this regression and that the volatility effect is stronger for bear LETFs than for bull LETFs. In this paper, we find a similar effect in that it is more difficult to track a negatively leveraged benchmark    than a positively leveraged one. 

Understanding the price dynamics and tracking performance of (L)ETFs are practically useful for developing trading strategies. For instance, \cite{TM} model the spread between mean-reverting pairs of gold and silver ETFs, and develop efficient algorithms for estimating the parameters of this model for trading purposes. \cite{GoldMinerSpreads2013} develop a    genetic programming algorithm   for trading gold ETFs. \cite{LeungLi} analyze the optimal sequential timing strategies for trading pairs of ETFs based on gold, gold miners, or silver.  Additionally,  \cite{Naylor} find gold and silver ETFs to be highly profitable ETFs and are able to yield highly abnormal returns based on filtering strategies.


%


\section{Gold Spot \&  Futures}\label{futures}
In this section we analyze the price dynamics of gold futures  with respect to the spot. One benchmark for the spot gold price are the London Gold Fixing Indices, GOLDLNAM and GOLDLNPM. Each of these indices is only updated once per business day:  10:30 AM for GOLDLNAM, and 3:00 PM for GOLDLNPM in London times, by four members of the London Bullion Market Association (Scotia-Mocatta, Barclays Bank, HSBC, and Societe-Generale).\footnote{According to the London Bullion Market Association (http://www.lbma.org.uk/pricing-and-statistics)}  Another widely used benchmark for the gold spot price is the Gold-U.S. Dollar exchange rate (XAU-USD). It indicates the U.S. dollar amount required to buy or sell one troy ounce of gold immediately. XAU-USD is frequently updated around the clock and its closing price is available for all trading days studied from 12/22/2008 through 7/14/2014. For these reasons, we choose XAU-USD as our benchmark for the gold spot price throughout this paper.

 In the gold futures market, the front months, such as the 1-month and 2-month contracts, are  actively traded daily. However,   other available  contracts are set to expire in  specific calendar  months   within the next few years. As such,  it may not always be possible to trade 6-month and 12-month futures, and one may need to alternate with the nearest month. For a 6-month futures position we assume a position which alternates between 6-month futures and 5-month futures and for a 12-month futures position we assume a position which alternates between 12-month futures and 11-month futures. This involves simply waiting while the 6-month futures (resp.  12-month) futures becomes a 5-month futures (resp.  11-month) after one month and then rolling the position forward 2 months after the second month passes. For example, if it now January 2012 a 12-month futures contract would be the Dec-12 contract. When February 2012 comes by, this becomes an 11 month contract, but the Jan-13 contract is not available. Instead, we hold the position as an 11-month futures and then in March 2012, we roll the position forward into the Feb-13 contract returning it to a 12-month position.  Throughout, we will use the 1, 2, 6, and 12 month gold futures contracts.

\subsection{Price Dependency}\label{futuresregression}
We begin by performing linear regressions of the 1-day returns of gold spot versus the futures of maturities: 1, 2, 6, and 12 months.  Across all maturities, the  linear relationships are all strong and they are  quite similar. In Table \ref{tab:Gold Regression Table}, we summarize  the regression results, including the slope, intercept, $R^2$, and root mean squared error (RMSE).  The $R^2$ values are all greater than 80\%, indicating a strong linear fit. For every maturity, the  slope is close to 0.94 and the intercept is essentially zero. The slopes suggest that the price sensitivity of futures with respect to the spot is slightly less than 1 to 1. While this may suggest  that the futures prices should be  be less volatile than the spot return, we find that the historical annual volatilities of the futures are higher: 19.261\% (1-month), 19.263\% (2-month), 19.269\% (6-month),  and 19.266\% (12-month), as compared to   the spot  (18.374\%). The fact that the regression results are almost the same among different futures suggests that the futures prices are highly positively correlated. Indeed, our calculations show that the correlation among the futures over the same period are all over 99\%.

\begin{table}[H]\centering\begin{small}

    \setlength{\extrarowheight}{2pt}
    \begin{tabular}{l r r r r} 
        \hline
        \text{Response} & \text{Slope} & \text{Intercept} & \text{$R^2$} & \text{$RMSE$} \\ \hline
        \hline
        1-Month                     & 0.94314                                   & $2.41916\cdot10^{-5}$     	  	& 0.80947\,         	 & 0.00530 \\
        2-Month                     & 0.94301                                    & $6.20316\cdot10^{-5}$ 	    	& 0.80911\,     	 	& 0.00530 \\
        6-Month                     & 0.94348                                   & $3.17973\cdot10^{-5}$ 	        	 	& 0.80934\,     		& 0.00530 \\
        12-Month                    & 0.94358                                     & $-5.23334\cdot10^{-5}$ 	 	& 0.80984\,       	& 0.00529 \\
       \hline    \end{tabular}\end{small}
   \parbox{0.8\textwidth}{\caption{\small{A summary of the regression coefficients and measures of goodness of fit for regressing one-day returns of 1, 2, 6 and 12-month futures on 1-day returns of spot gold from  12/22/2008 to 7/14/2014.}}\label{tab:Gold Regression Table}}    
\end{table}



\begin{figure}[H]
\centering
    \includegraphics[width=6in]{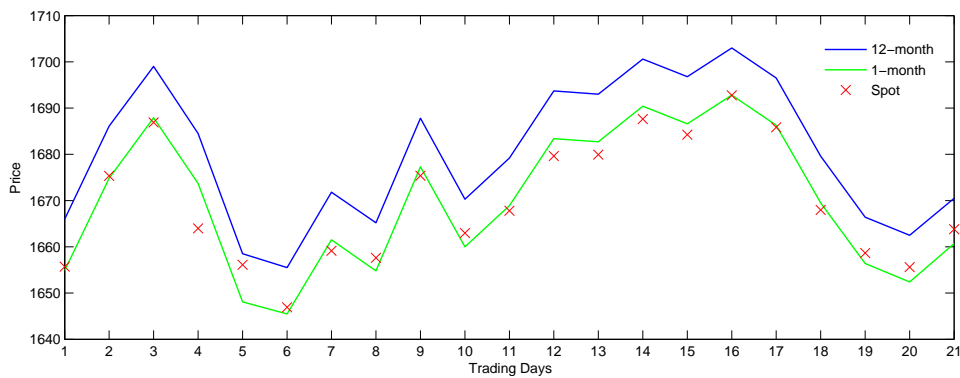}
    \parbox{0.8\textwidth}{\caption{\small{Time evolution of spot gold price, 1-month futures price (Jan-13 contract) and 12 month futures price (Dec-13 contract) over the period from Dec 29, 2012 to Jan 29, 2013. We can see the two futures prices move in parallel, and that the spot price trades very close to the front month futures price.}}\label{fig:correlationfigure}}
\end{figure}

The high correlation among futures prices can also be seen from their time series.  In Figure \ref{fig:correlationfigure}, we plot the gold price, 1-month futures price (Jan-13 contract) and 12 month futures price (Dec-13 contract) over the period from 12/29/2012 to 1/29/2013. Over this period the 1-month futures and 12-month futures prices move in parallel to one another just as the near perfect correlation would indicate. Furthermore, the gold spot price and 1-month futures price are close, as expected. On Jan 29, 2013, or trading day 21 in Figure \ref{fig:correlationfigure}, the 1-month futures would expire. We observe a slight discrepancy between the futures price and the spot on this date. While  futures should theoretically converge to the spot price at maturity,  in practice gold futures settle at their volume weighted average (VWAP) price within the last one minute.\footnote{According to CME Group gold futures settlement procedures documentation, available at http://www.cmegroup.com/trading/metals/files/daily-settlement-procedure-gold-futures.pdf} The last price will be equal to the spot price at maturity, but the settlement price used here will be calculated by weighting this last price against its volume traded and hence need not be equal to the spot gold price.

\begin{figure}
    \begin{centering}
    \subfigure[Jan-Jun 2009]{\includegraphics[width=3in]{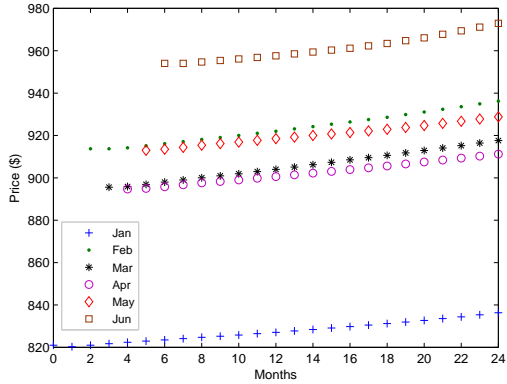}}
    \subfigure[Jan-Jun 2013]{\includegraphics[width=3in]{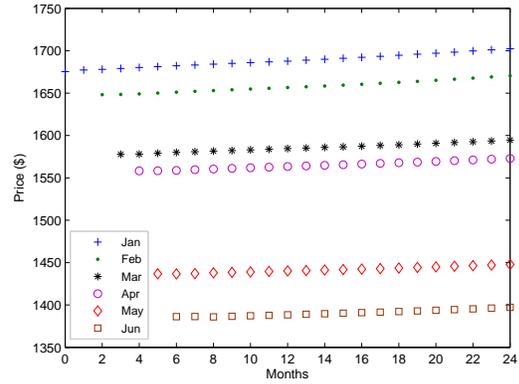}}
    \caption{\small{Term structures from Jan to Jun in 2009 (left) and 2013 (right). }}
    \label{fig:termstructure}
    \end{centering}
\end{figure}

In fact, the price co-movement among futures also has implications for the term structure dynamics. On a typical date in the gold market, futures prices are increasing and convex with respect to maturity as seen in Figure \ref{fig:termstructure}. Since the futures contracts tend to move in parallel, this leads to almost parallel shifts in the term structure. The shape of the term structure remains almost the same over time. During both periods we can see the increasing convex feature of the gold futures market, but in 2013 there is a reduction in convexity. In 2009, the term structure generally shifts upward from Jan to Jun, while in 2013 the term structure strictly shifts downward from Jan to Jun.

\begin{figure}[H]
\centering
    \subfigure[1-Day Returns]{\includegraphics[width=3in]{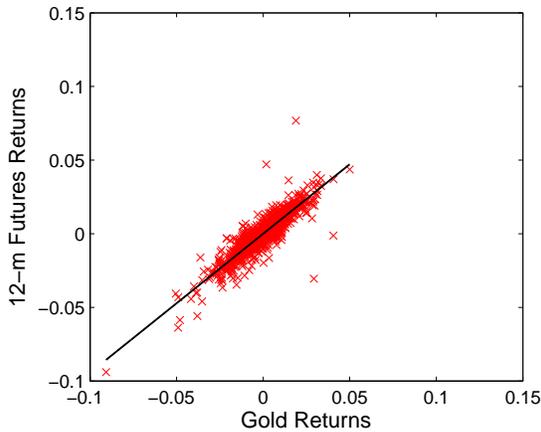}}
    \subfigure[10-Day Returns]{\includegraphics[width=3in]{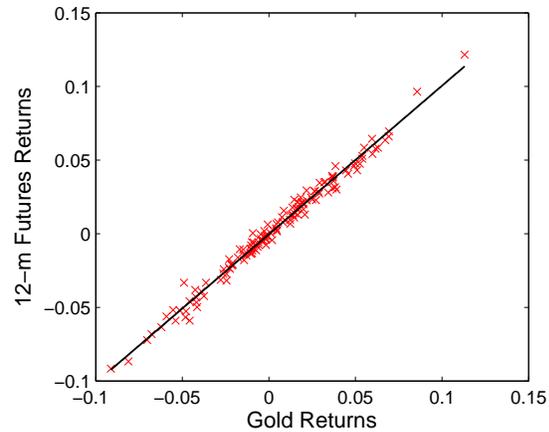}}
   \parbox{0.8\textwidth}{\caption{\small{Linear regressions of 12-month futures returns based on 1-day returns (left) and 10-day returns (right) versus the  spot gold returns.}}\label{fig:Gold Regression Futures}}
\end{figure}

\newpage
Next, we compare the linear relationships for returns computed over  5, 10, and 15 trading days. Since we use disjoint time windows for return calculations, a longer holding period implies fewer data points for the regression. In Figure \ref{fig:Gold Regression Futures}, we plot the regressions of 12-month futures returns versus gold returns for both 1-day returns and 10-day returns, plotted on the same $x$-$y$ axis scale. We can see the returns vary on a larger range for the 10-day returns. Spot gold has a 1-day return between -9.07\% (4/15/2014)\footnote{See http://mobile.nytimes.com/blogs/dealbook/2013/04/15/golds-plunge-shakes-confidence-in-a-haven/} and 4.99\% (1/23/2009), while its 10-day returns vary between -9.12\% (6/6/2013 to 6/19/2013) and 11.30\% (8/5/2011 to 8/18/2011). On the other hand, the 12-month futures has a 1-day return between -9.40\% (4/15/2014) and 7.68\% (3/19/2009), while its 10-day returns vary between -9.16\% (6/6/2013 to 6/19/2013) and 11.30\% (8/5/2011 to 8/18/2011).  Moreover, we can see that the slope is slightly higher for the 10-day returns versus the 1-day returns. 

This is confirmed numerically and in general for the various futures contracts in Table \ref{tab:holdingperiodslopes}. Here, we give the slopes for the regression of each futures return versus the gold return, while varying the holding period. We display the slopes and $R^2$ values from Table \ref{tab:Gold Regression Table} for comparison. However, we do not give the intercepts for these regressions as they are all very trivial and effectively 0. We can see that the slopes approach the value 1 as the holding period is lengthened. Thus, the longer the holding period, the more closely the gold return and futures price return are to one another. In particular, the slopes for 10-day returns are all greater than 1, indicating an increased price sensitivity. Furthermore, the strength of this linear relationship increases as can be seen by the increasing $R^2$ values in Table \ref{tab:holdingperiodslopes}.\\

\begin{table}[H]\centering \begin{small}
    \setlength{\extrarowheight}{2pt}
    \begin{tabular}{l l r r r r }
        \hline
        &\text{Days} & \text{1-Month} & \text{2-Month} & \text{6-Month} & \text{12-Month} \\ \hline
        \hline   
     {Slope}   & 1                     & 0.94314                                   & 0.94301    	  	& 0.94348    	 & 0.94358 \\
        		   & 5                     & 0.99805                                   & 0.99752    	  	& 0.99757    	 & 0.99715 \\
        		   & 10                   & 1.01075                                   & 1.01020    	  	& 1.00970    	 & 1.00841 \\
        		   & 15                   & 0.99448                                   & 0.99461    	  	& 0.99431    	 & 0.99345 \\
              \hline \hline
    \text{$R^2$}   & 1                     & 0.80947                                   & 0.80911    	  	& 0.80934    	 & 0.80984 \\
              	   	    & 5                     & 0.97154                                   & 0.97158    	  	& 0.97194    	 & 0.97165 \\
              	   	    & 10                   & 0.98516                                   & 0.98530    	  	& 0.98572    	 & 0.98550 \\
              	   	    & 15                   & 0.98704                                   & 0.98691    	  	& 0.98725    	 & 0.98713 \\
              \hline    \end{tabular}
            \end{small}     \parbox{0.8\textwidth}{ \caption{\small{A summary of the slopes and $R^2$ from the regressions of futures returns versus gold returns over different holding periods.}}\label{tab:holdingperiodslopes}}
    \end{table}

\subsection{Static Replication of Gold Spot Price}\label{staticrepfutures}
In this section we consider replication of the gold spot price with a static portfolio of futures contracts. We use portfolios of either 1 or 2 futures contracts and an investment in the money market account. We seek a static portfolio that  minimizes the sum of squared errors:
\begin{equation}\label{ssedeffutures}
SSE=\sum_{j=1}^n(V_j-G_j)^2\,,
\end{equation}
where $V_j$ is the dollar value of the portfolio on trading day $j$, while $G_j$ is the dollar value of the gold spot price on trading day $j$. 

Let $k$ be the number of futures contracts and $\textbf{w} := (w_0, \ldots, w_{k})$ be the real-valued vector of portfolio weights. In particular,  $w_{0}$ represents the weight given to the money market account. To calculate the optimal portfolio value we will choose weights historically which minimize SSE over the 5-year period 12/22/2008 through 12/22/2013. Thus, we solve the following constrained least squares optimization problem:

\begin{equation}
\begin{aligned}
& \underset{\textbf{w} \in \R^{k+1}}{\text{min}}
& & \| \textbf{C}\textbf{w}-\textbf{d}\|^2 \\
& \text{s.t.}
& & \sum_{j=0}^{k} w_j=1
\end{aligned}
\end{equation}

The matrix $\textbf{C}$ contains as columns, the historical prices of the various futures contracts and the money market account,\footnote{We use historical overnight LIBOR to construct an investment in the money market account.} and the vector, $\textbf{d}$ contains the historical prices of spot gold. These prices are normalized by $\$1000$, without loss of generality,  so that an investor starting with $\$1000$ will invest \$$1000\cdot w_j$ into the $j$th futures contract and \$ $1000\cdot w_0$ into the money market account.  

We will compare our portfolios to investments in the ETF GLD, which tracks the gold spot price. To do this, we will perform an out-of-sample analysis and compare the values of $\$1000$ invested in GLD and $\$1000$ invested in our constructed portfolio over the period from 12/23/2013 to 7/14/2014. To measure  the performance we will use  the following root mean squared error for  both in-sample and out-of-sample prices:
\begin{equation}\label{rmsecalculation}
RMSE =  \sqrt{\frac{1}{n} \sum_{j=1}^n \left({V_j-G_j}\right)^2}\,.
\end{equation}

 We solve this optimization problem for ten different portfolios with  1  or 2 futures, along with the  money market account.  The optimal weights, and corresponding in-sample/out-of-sample RMSEs are given in Table \ref{tab:Static Replication Table}. In general, the money market account is  used minimally as the weights  on the account are less than 7\% in absolute value  for all ten portfolios.  

For all the portfolios with 2 futures contracts, the optimal strategy  is to go long the shorter term futures contract and  short the longer term futures contract, with different weights. The  sum of the two resulting weights are approximately 1.  In terms of RMSE,  the 1-month futures contract appears to be the best replicating instrument  of the gold spot. When it is used alone,  it performs best relative to other single futures portfolios. When it is used in a pair with another futures,  it performs better than any other single futures contract, and better than all other  futures pairs:  (2-m, 6-m), (2-m, 12-m), and (6-m, 12-m).

\begin{table}[h]\centering\begin{small}
    \setlength{\extrarowheight}{2pt}
    \begin{tabular}{l l r r r r r } 
        \hline
        & \text{Futures} & \text{$w_0$} & \text{$w_1$} & \text{$w_2$} & \text{$RMSE$} (in) & \text{$RMSE$} (out)\\ 
        \hline
        \hline
        {1 Futures} & 1-m                     & -0.01071 	& 1.01071      	  	& -          			& 6.62989   	& 3.34047 \\
        			 & 2-m                     & -0.04835 		& 1.04835     	  	& -          			& 12.22148 	& 6.12074 \\
        			 & 6-m                     & -0.05030		& 1.05030      	  	& -          			& 13.23663 	& 5.33971 \\
        			 & 12-m                   & -0.06842		& 1.06842     	  	& -          			& 15.11103 	& 5.71318 \\
        \hline
        \hline
        {2 Futures} & 1-m, 2-m               & -0.00088	& 1.27315                  & -0.27227         	& 6.26711 	& 2.79411 \\
        			 & 1-m, 6-m               & -0.00171	& 1.23899                  & -0.23729        	& 6.27232 	& 2.97602 \\
        			 & 1-m, 12-m             & -0.00021	& 1.19336	                  & -0.19315         	& 6.28735 	& 3.00006 \\
        			 & 2-m, 6-m               & -0.04079	& 5.06602                  & -4.02523       	& 10.27413 	& 9.37292 \\
        			 & 2-m, 12-m             & -0.01179	& 2.94860                  & -1.93681  	   	& 9.65705 	& 7.08414 \\
        			 & 6-m, 12-m             & 0.01481		& 4.80979                  & -3.82460     		& 9.57938 	& 4.52846 \\
       \hline    \end{tabular}
   \parbox{0.8\textwidth}{\caption{\small{A summary of the weights and in/out of sample RMSEs for portfolios of 1 and 2 futures contracts. For portfolios with 2 futures, the weight on the shorter term futures  is $w_1$. For portfolios with a single futures, we have $w_2=0$.  The weight assigned to the money market account is denoted by $w_0$.}}\label{tab:Static Replication Table}}
    \end{small}
\end{table}

\begin{figure}[h] \centering
	\subfigure[Portfolio with 1-m Futures]{\includegraphics[width=6in]{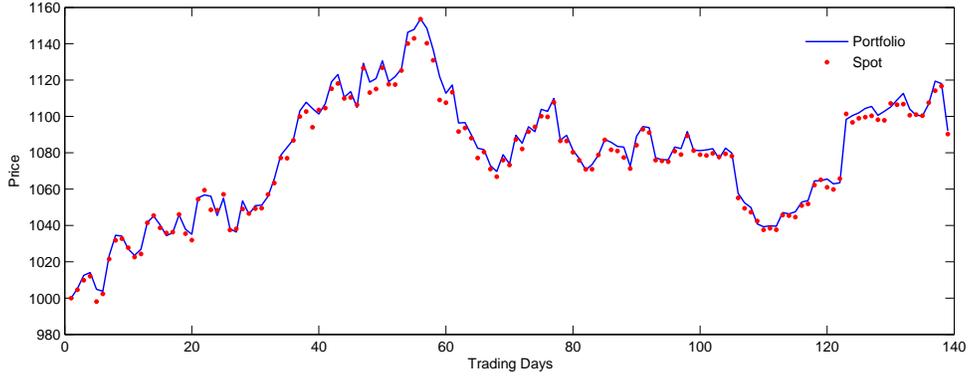}}
	\subfigure[GLD]{\includegraphics[width=6in]{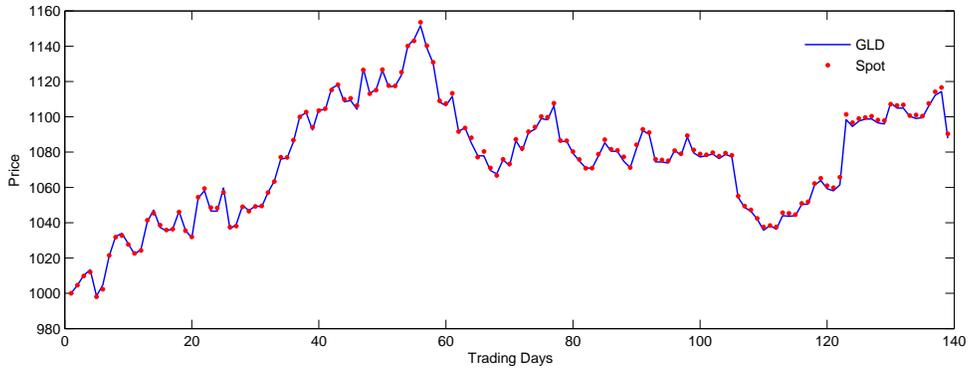}}
	\parbox{0.85\textwidth}{\caption{\small{Out-of-sample time series of our optimal portfolio of front month futures and money market account (top) compared to the spot price,  and GLD compared to the spot price (bottom).  The trading days are over the period 12/23/2013 to 7/14/2014.}}\label{fig:outofsampleFutures}}
\end{figure}

In the sample, the RMSE values for the futures portfolios range from 6.63 to 15.11. Since these values are based on a $\$1000$ investment, this means the error within the sample is between 0.663\% and 1.511\%, which is quite low. By comparison, our calculations give  a RMSE of 2.091\% for the gold ETF  (GLD) during  12/22/2008 to  12/22/2013. Over this longer horizon of 5 years, our portfolios track the benchmark better than GLD. However, over the more recent, shorter out-of-sample period, 12/23/2013 to 7/14/2014, GLD  appears to track spot gold slightly  better. The RMSE value for GLD during this period is 0.128\%, whereas our best portfolio gives a RMSE of 0.279\%.  In    Figure \ref{fig:outofsampleFutures}, we show the time series of the optimal static portfolio with the front month futures (top), and the time series for GLD. It is visible that both  track the gold spot price closely over this  out-of-sample period.

\newpage

\section{Leveraged ETFs}\label{LETFs}
In this section we analyze  the returns and tracking performances of various leveraged ETFs. From historical  prices of  each LETF, we conduct an estimation of the leverage ratio, and investigate the potential deviation from the target leverage ratio. Moreover, we construct a number of   static  portfolios   with futures contracts to seek replication of some  leveraged benchmarks. However, the  static portfolios fail to effectively track the  leveraged benchmarks. This motivates us to  consider a dynamic portfolio with futures, which turns out to have a much better tracking performance. 

By design, an LETF seeks to provide  a constant multiple of the daily returns of an underlying index or asset. Let us denote $\beta \in \{-3, -2, +2, +3\}$  the   leverage ratio stated by the LETF, and $R_j$ the daily return of the underlying (gold spot). Ideally, the LETF value on day $n$, denoted by $L_n$, is given by 
\begin{equation}\label{leveredindexcreation}
L_n=L_0 \cdot \prod_{j=1}^n (1+\beta \, R_j).
\end{equation}
We call this  the leveraged benchmark, and examine the empirical  performance of various  LETFs with respect to this benchmark. 

For many  investors, one  appeal of LETFs is that leverage can amplify returns when the underlying is moving in the desired direction. Mathematically,  we  can see this  as follows. Rearranging \eqref{leveredindexcreation} and taking the derivative of the logarithm, we have
\begin{equation}\label{logindex}
\frac{d}{d\beta}\left(\log \left(\frac{L_n}{L_0}\right)\right)= \sum_{j=1}^n \frac{R_j}{ 1+\beta \, R_j}.
\end{equation}
With a positive leverage ratio  $\beta>0$, if $R_j>0$ for all $j$, then    $\log\left(\frac{L_n}{L_0}\right)$, or equivalently the value $L_n$, is increasing in $\beta$. In other words,  when the underlying asset is increasing in value,  a larger, positive leverage ratio  is preferred.  On the other hand, if $R_j<0$ for all $j$, and  $\beta<0$, a more negative $\beta$ increases $\log\left(\frac{L_n}{L_0}\right)$ and thus  $L_n$. This means that when the underlying asset is decreasing in value,  a more negative leverage ratio yields a higher return.

The example below illustrates the consequences of maintaining a constant leverage in an environment with non-directional movements:

\begin{center}
	\begin{tabular}{c c c c c c c }
		\hline
		\text{Day} & \text{ETF} & \%-change & \text{+2x LETF} & \%-change  & \text{$-2$x LETF} & \%-change  \\
		\hline
		\hline
		0 & 100 &  & 100 &  & 100 &  \\
		1 & 98 & -2\% & 96 & -4\% & 104 & 4\% \\
		2 & 99.96 & 2\% & 99.84 & 4\% & 99.84 & -4\% \\
		3 & 97.96 & -2\% & 95.85 & -4\% & 103.83 & 4\% \\
		4 & 99.92 & 2\% & 99.68 & 4\% & 99.68 & -4\% \\
		5 & 97.92 & -2\% & 95.69 & -4\% & 103.67 & 4\% \\
		6 & 99.88 & 2\% & 99.52 & 4\% & 99.52 & -4\%\\
		\hline
	\end{tabular}
\end{center}

Even though the ETF records a tiny loss of 0.12\% after 6 days, the +2x LETF ends up with a loss of 0.48\%, which is greater (in absolute value) than 2 times the return ($-0.12\%$) of the ETF. We can see this to be the case on any day (e.g. not just the terminal date) except for day 1. For example, on day 3, the ETF has a net loss of 2.04\% and the LETF has a net loss of 4.15\%, which is greater (in absolute value) than 4.08\% (twice the absolute value of the return of the ETF). Furthermore, it might be intuitive that 	that the $-2$x LETF should have a positive return when the ETF and LETF have negative returns, this is not true. At the terminal date, both the long and short LETFs have recorded net losses of 0.48\%. Again, this occurs throughout the period as well, not just the terminal date. In addition to day 6, both the long and short LETFs as well as the ETF itself are in the black. These results are consequences of volatility decay.

\begin{figure}[H]
\centering
    \includegraphics[width=6in]{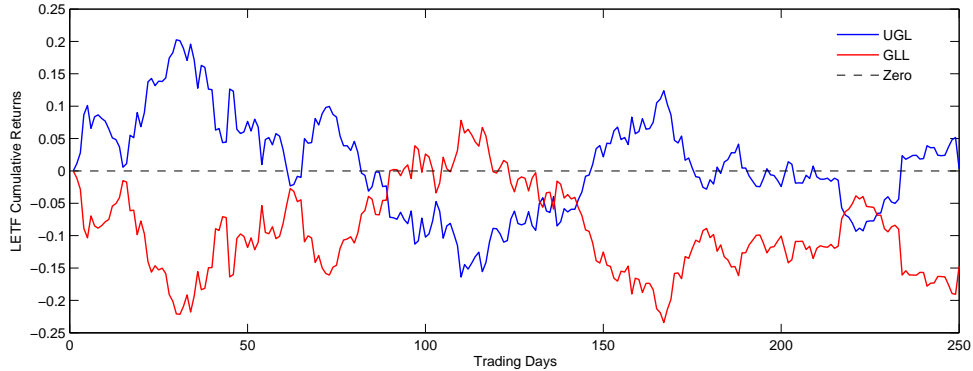}
    \parbox{0.8\textwidth}{\caption{\small{UGL (+2x) and GLL ($-2$x) cumulative returns from July 2013 to July 2014. Observe that both UGL and GLL can give negative returns (below the dotted line of 0\%) simultaneously over several periods in time.}}\label{fig:letfcumulativereturns}}
\end{figure}

Although long and short LETFs are expected to move in opposite directions daily by design, it is often possible for both LETFs to have negative cumulative returns when held over a longer horizon. Figure \ref{fig:letfcumulativereturns} shows the historical cumulative returns of the gold LETFs UGL (+2x) and GLL ($-2$x) from July 2013 to July 2014. From trading day 124 (1/24/2014) onward, GLL has a negative cumulative return. There are points after trading date 124 where UGL also has a negative cumulative return. In fact, it starts in the black on this date and continues to have a net loss until trading date 146 (2/12/2014). This occurs again a few times, another long stretch where both have a net loss is trading date 210 (5/15/2014) through 233 (6/18/2014).This observation, though maybe counter-intuitive at first glance, is a consequence of daily replication of leveraged returns. The value erosion tends to accelerate during periods of non-directional movements.

\subsection{Empirical Leverage Estimation}\label{letfregression}
We conduct a regression analysis and the results are given in Table \ref{tab:LETF Regression Table}. Each slope is approximately equal to the LETF's target leverage ratio. In principle, if each (L)ETF is able to generate the desired multiple of daily returns, the slopes of the regression should be equal to the various leverage ratios. In this table, we give an additional two columns for the t-statistic and p-value for testing the hypothesis: $\{H_0:\, \text{slope}= \beta \}$ vs. $\{H_1:\, \text{slope} \neq \beta\}$. Here, $\beta$ is the target leverage ratio. We can see that each p-value is larger than 0.05 and therefore conclude that statistically, each (L)ETF does not differ from its target leverage ratio. This demonstrates to us that the (L)ETFs are performing exactly as desired, at least on a daily basis. 

\begin{table}[H]\centering\begin{small}
    \setlength{\extrarowheight}{3pt}
    \begin{tabular}{l r r r  r r r } 
        \hline
        \text{(L)ETF} & \text{Slope} & \text{Intercept} & \text{t-stat} & \text{p-value}& \text{$R^2$} & \text{$RMSE$} \\ \hline
        \hline
        GLD                      & 1.00540                   & $-1.64276\cdot10^{-5}$ 	& 1.42692\         & 0.15382    & 0.98060\      	  & 0.00163 \\
        UGL                      & 2.00572                   & $-1.31073\cdot10^{-4}$   	& 0.73319\         & 0.46357    & 0.97930\             & 0.00337 \\
        GLL                       & -2.00556                  & $-1.05504\cdot10^{-4}$    & 0.67089\     	 & 0.50240    & 0.97673\     	  & 0.00358 \\
        UGLD                    & 2.99358                  & $-1.98605\cdot10^{-4}$   	& 0.45211\   	 & 0.65133    & 0.98484\     	  & 0.00421 \\
        DGLD                    & -2.97528                 & $-1.69127\cdot10^{-5}$  	& 0.97442\         & 0.33019    & 0.95256\     	  & 0.00752\\
       \hline    \end{tabular}\end{small}
  \parbox{0.8\textwidth}{ \caption{\small{A summary of the regression coefficients and measures of goodness of fit for regressing one-day returns of (L)ETFs versus spot gold. We include 2 additional columns for the t-statistic and p-value for testing the hypothesis that the slope equals the leverage ratio in each case.}}\label{tab:LETF Regression Table}}
\end{table}

We see also that the $R^2$ values for each regression are quite high, all above 95\%. Next, we compare the long and short LETFs for a fixed $|\beta|\in\{2, 3\}$. The short LETF tends to have a higher RMSE and lower $R^2$ value. Finally, we see in general that as the leverage ratio increases in absolute value, there is a higher RMSE. One possible explanation is that the benchmark is leveraged, and this could magnify the tracking error.

Just as in Section \ref{futuresregression}, we analyze the effects of changing the holding period. In Table \ref{tab:letfholdingperiod} we give the slopes and intercepts for the regressions of each (L)ETF's return versus the spot return while varying the holding period between 1 and 5 days. Our computations show the $R^2$ values are all above 95\%. We can see that the slopes all approximately equal the target leverage ratio of the (L)ETF. However we notice that in general the intercepts get more negative as the holding period is lengthened. Although they are still quite small, they become more significant as the holding period increases. Our calculations show that the p-values for testing the hypothesis: $\{H_0:\, \text{intercept}= 0 \}$ vs. $\{H_1:\, \text{intercept} \neq 0\}$ generally tend to decrease for each (L)ETF. In fact, for UGL the intercepts turn out to be statistically different from 0 (at the 5\% level) for holding periods of 3, 4 and 5 days with p-values of 1.37\%, 0.57\% and 0.33\%, respectively. This is consistent with the volatility decay discussed above. We saw there an example where over shorter periods, the LETF tracks its leverage ratio well, but over a longer period it tends to lose money when there is high volatility. The intercepts being different from 0 is akin to the volatility decay in the following sense. Over longer periods, the regressions show that we require more information than just the gold return to predict the LETF return. 

To compare the performance of the LETF versus the target multiple of the spot return, we also report in Table \ref{tab:letfholdingperiod}, the average  return differential, defined by 
\begin{equation}\label{mtecalculation}
\overline{RD}=\frac{1}{m}  \sum_{j=1}^m \left(R^{(L)}_j - \beta \cdot R^{(G)}_j\right),
\end{equation}
where $m$ is the number of the periods, $R^{(L)}_j$ is the LETF's return over the holding period and $R^{(G)}_j$ is the spot's return over the holding period. We find this to be increasing (in absolute value) with the holding period length. That is, as we hold the LETF longer, it tends to increasingly underperform with respect to the multiple of the underlying return, on average. This is exactly the same notion described above, since over time, the volatility of the underlying causes the LETF to erode in value. 

\begin{table}[H]\centering\begin{small}
    \setlength{\extrarowheight}{2pt}
    \begin{tabular}{l c r r r r r}
        \hline
         & \text{Days} & \text{UGL} & \text{GLL} & \text{UGLD} & \text{DGLD} & \text{GLD} \\
   	\hline
   	\hline
   {Slope} & 1   	& 2.00572  & -2.00556   & 2.99160  & -2.96362   & 1.00540 \\
   	       & 2 	& 2.00828  & -2.00395   & 2.92015  & -3.02429   & 1.00477 \\
               & 3 	& 1.97770  & -1.99661   & 2.94520  & -3.06690   & 0.99194 \\
   	       & 4	& 2.00071  & -2.00908   & 2.97854  & -3.04311   & 1.00206 \\
   	       & 5	& 2.02081  & -2.03273   & 2.89478  & -3.07267   & 1.01039 \\
   \hline
   \hline
    {Intercept $(\cdot10^{-4})$} & 1 & -1.31074    & -1.05505    & -1.98605    & -0.16913    & -0.16428 \\
    		     			    & 2 & -2.73282    & -2.28807    & -4.23120    & -1.81545    & -0.33388 \\
    		     			    & 3 & -4.06417    & -3.96629    & -4.97993    & -1.05631    & -0.40843 \\
    		     			    & 4 & -5.44427    & -4.62166    & -8.08478    & -2.69971    & -0.66133 \\
    		     			    & 5 & -6.81614    & -4.82805    & -8.03417    & -0.30766    & -0.92359 \\
	\hline
	\hline				    
     {$\overline{RD}\, (\cdot10^{-3})$} & 1 & -0.12892  & -0.10760  & -0.19673  & -0.02414  & -0.01439 \\
  				& 2 & -0.26677  & -0.23191  & -0.43999  & -0.19081  & -0.02964 \\
  				& 3 & -0.43253  & -0.39265  & -0.53511  & -0.08462  & -0.05028 \\
  				& 4 & -0.54331  & -0.47645  & -0.81687  & -0.27854  & -0.06290 \\
  				& 5 & -0.64076  & -0.54705  & -0.79183  &  0.07165  & -0.07195 \\					    
			    
              \hline    \end{tabular}\end{small}
  \parbox{0.8\textwidth}{ \caption{\small{A summary of the slopes and  intercepts from the regressions of LETF returns versus gold returns, as well as the average return differential ($\overline{RD}$) over different holding periods.}}\label{tab:letfholdingperiod}}
\end{table}


\subsection{Static Leverage Replication}\label{staticrepletfs}
In this section, we perform the same optimization as in Section \ref{staticrepfutures}. We once again seek a static portfolio of futures which minimizes SSE. Let $k$ be the number of futures contracts and $\textbf{w} := (w_0, \ldots, w_{k})$ be the real-valued vector of portfolio weights. As before, $w_{0}$ represents the weight given to the money market account. We seek the weights which minimize SSE over the 5-year period 12/22/2008 through 12/22/2013. Thus, we are led to the same constrained least squares optimization problem:

\begin{equation}
\begin{aligned}
& \underset{\textbf{w} \in \R^{k+1}}{\text{min}}
& & \| \textbf{C}\textbf{w}-\textbf{L}\|^2 \\
& \text{s.t.}
& & \sum_{j=0}^{k} w_j=1
\end{aligned}
\end{equation}

Again, the matrix $\textbf{C}$ contains as columns, the historical prices of the various futures contracts and the money market account. Here, the vector $\textbf{L}$ contains the historical prices of the leveraged  benchmark in   \eqref{leveredindexcreation}. Without loss of generality, we normalize the prices by $\$1000$ so that our solution will give us a set of weights on each instrument. 


We will compare the tracking error of our optimal portfolios to that of investments in the LETFs. To be able to analyze the portfolios we get by solving the optimization problem, we will perform an out of sample analysis over the period 12/23/2013 through 7/14/2014 and see how $\$1000$ invested in the LETFs and $\$1000$ invested in our optimal portfolios perform. To quantify the performance we use the same root mean square error 

\begin{equation}\label{rmsecalculationletfs}
RMSE =  \sqrt{\frac{1}{n} \sum_{j=1}^n \left({V_j-L_j}\right)^2},
\end{equation}
where, $V_j$ is the dollar value of the portfolio on trading day $j$, while $L_j$ is the dollar value of the leveraged benchmark on trading day $j$. Now, we present the results for the optimization and in sample/out of sample RMSE.

\begin{table}[H]\centering\begin{small}

    \setlength{\extrarowheight}{2pt}
    \begin{tabular}{l l r r r r r } 
        \hline
        \text{UGL(+2x)} & \text{Futures} & \text{$w_0$} & \text{$w_1$} & \text{$w_2$} & \text{$RMSE$} (in) & \text{$RMSE$} (out)\\ 
        \hline
        \hline
        {1 Futures} & 1-m                     & -1.48159 	& 2.48159      	  	& -          			& 153.20152   	& 41.48530 \\
        			 & 2-m                     & -1.57601 	& 2.57601     	  	& -          			& 140.41254 	& 49.56627 \\
        			 & 6-m                     & -1.58107	& 2.58107      	  	& -          			& 138.63116 	& 47.45953 \\
        			 & 12-m                   & -1.62627	& 2.62627     	  	& -          			& 134.50683 	& 48.29851 \\
        \hline
        \hline
        {2 Futures} & 1-m, 2-m               & -1.94496	& -9.89441                  & 12.83937         	& 114.30949 	& 81.36970 \\
        			 & 1-m, 6-m               & -1.91184	& -8.43451                  & 11.34634        	& 113.67580 	& 67.39428 \\
        			 & 1-m, 12-m             & -2.02222	& -6.92791                  & 9.95013         	& 108.29897 	& 67.15776 \\
        			 & 2-m, 6-m               & -1.64536	& -34.26195                & 36.90731       	& 126.62074	& 22.64952 \\
        			 & 2-m, 12-m             & -1.99096	& -18.99151                & 21.98248  	   	& 111.75046	& 40.92852 \\
        			 & 6-m, 12-m             & -2.24344	& -35.66917                & 38.91260     	& 102.86283 	& 62.53830 \\
       \hline    \end{tabular}
   \parbox{0.8\textwidth}{\caption[H]{\small{A summary of the weights and in/out of sample RMSEs for portfolios of 1 and 2 futures contracts which attempt to replicate a leveraged benchmark with $\beta=2$. By comparison, the +2x LETF, UGL has an out of sample RMSE of only 5.52485.}}\label{tab:Static Replication Table UGL}}
	\end{small}
\end{table}

\begin{table}[H]\centering\begin{small}

    \setlength{\extrarowheight}{2pt}
    \begin{tabular}{l l r r r r r } 
        \hline
        \text{GLL($-2$x)} & \text{Futures} & \text{$w_0$} & \text{$w_1$} & \text{$w_2$} & \text{$RMSE$} (in) & \text{$RMSE$} (out)\\ 
        \hline
        \hline
        {1 Futures} & 1-m                     & 1.97544 	& -0.97544     	  	& -          			& 152.33349   	& 76.14381 \\
        			 & 2-m                     & 2.01068	& -1.01068     	  	& -          			& 155.76147 	& 73.14698 \\
        			 & 6-m                     & 2.01234	& -1.01234      	  	& -          			& 156.43931 	& 74.00595 \\
        			 & 12-m                   & 2.02926 	& -1.02926     	  	& -          			& 158.06790 	& 73.79237 \\
        \hline
        \hline
        {2 Futures} & 1-m, 2-m               & 1.69510	& -8.46293                  & 7.76783         	& 139.27436 	& 100.10091 \\
        			 & 1-m, 6-m               & 1.70510	& -7.83463                  & 7.12954        	& 137.98773 	& 92.21120   \\
        			 & 1-m, 12-m             & 1.60773	& -7.37540                  & 6.76767         	& 133.31705 	& 93.11206 \\
        			 & 2-m, 6-m               & 1.93900	& -39.08635                & 38.14735       	& 142.57324 	& 43.40713 \\
        			 & 2-m, 12-m             & 1.57681	& -23.56157                & 22.98477  	   	& 127.90630	& 62.10415 \\
        			 & 6-m, 12-m             & 1.27887	& -43.36906                & -43.09020     	& 117.81839 	& 88.35987 \\
       \hline    \end{tabular}
   \parbox{0.8\textwidth}{\caption[H]{\small{A summary of the weights and in/out of sample RMSEs for portfolios of 1 and 2 futures contracts which attempt to replicate a leveraged benchmark with $\beta=-2$. By comparison, the $-2$x LETF, GLL has an out of sample RMSE of only 4.76269.}}\label{tab:Static Replication Table GLL}}
    \end{small}
\end{table}

\begin{table}[H]\centering\begin{small}
    \setlength{\extrarowheight}{2pt}
    \begin{tabular}{l l r r r r r } 
        \hline
        \text{UGLD(+3x)} & \text{Futures} & \text{$w_0$} & \text{$w_1$} & \text{$w_2$} & \text{$RMSE$} (in) & \text{$RMSE$} (out)\\ 
        \hline
        \hline
        {1 Futures} & 1-m                     & -3.33704 	& 4.33704      	  	& -          			& 555.49151   & 111.51133 \\
        			 & 2-m                     & -3.50625 	& 4.50625     	  	& -          			& 529.90858   & 125.98635 \\
        			 & 6-m                     & -3.51562	& 4.51562      	  	& -          			& 526.58448   & 122.33175 \\
        			 & 12-m                   & -3.59587	& 4.59587     	  	& -          			& 518.96896   & 123.83019 \\
        \hline
        \hline
        {2 Futures} & 1-m, 2-m               & -5.18147	& -44.92498                  & 51.10645         	& 379.11676 	& 270.58292 \\
        			 & 1-m, 6-m               & -5.02680	& -38.53507                  & 44.56188        	& 381.92109 	& 213.62168 \\
        			 & 1-m, 12-m             & -5.41970	& -31.91076	            & 38.33046         	& 366.49764 	& 210.91245 \\
        			 & 2-m, 6-m               & -3.78077	& -141.31534                & 146.09611       	& 472.32904 	& 48.04126   \\
        			 & 2-m, 12-m             & -5.12561	& -79.66101                  & 85.78662  	   	& 413.19682 	& 98.33728   \\
        			 & 6-m, 12-m             & -6.14011	& -147.04408                & 154.18419    	& 376.40048 	& 185.87170 \\
       \hline    \end{tabular}
  \parbox{0.8\textwidth}{ \caption{\small{A summary of the weights and in/out of sample RMSEs for portfolios of 1 and 2 futures contracts which attempt to replicate a leveraged benchmark with $\beta=3$. By comparison, the +3x LETF, UGLD has an out of sample RMSE of only 6.08133.}}\label{tab:Static Replication Table UGLD}}
    \end{small}
\end{table}

\begin{table}[H]\centering\begin{small}

    \setlength{\extrarowheight}{2pt}
    \begin{tabular}{l l r r r r r } 
        \hline
        \text{DGLD($-3$x)}& \text{Futures} & \text{$w_0$} & \text{$w_1$} & \text{$w_2$} & \text{$RMSE$} (in) & \text{$RMSE$} (out)\\ 
        \hline
        \hline
        {1 Futures} & 1-m                     & 2.14737 	& -1.14737      	  	& -          			& 222.59075   	& 135.67755 \\
        			 & 2-m                     & 2.18885 	& -1.18885     	  	& -          			& 225.76707 	& 132.14962 \\
        			 & 6-m                     & 2.19079	& -1.19079      	  	& -          			& 226.42907 	& 133.16229 \\
        			 & 12-m                   & 2.21063	& -1.21063     	  	& -          			& 228.13706 	& 132.91661 \\
        \hline
        \hline
        {2 Futures} & 1-m, 2-m               & 1.82654	& -9.71612                  & 8.88958         	& 211.09051 	& 163.04734 \\
        			 & 1-m, 6-m               & 1.83531	& -9.06486                  & 8.22955        	& 209.75562 	& 154.16271 \\
        			 & 1-m, 12-m             & 1.70781	& -8.79773	          & 8.08992         	& 204.41288 	& 155.82542 \\
        			 & 2-m, 6-m               & 2.10262	& -46.99103                & 45.88841       	& 212.78493 	& 95.24835 \\
        			 & 2-m, 12-m             & 1.63980	& -29.72604                & 29.08624  	   	& 195.74796 	& 117.35419 \\
        			 & 6-m, 12-m             & 1.24461	& -55.83130                & 55.58669     	& 183.42198 	& 150.58261 \\
       \hline    \end{tabular}
  \parbox{0.8\textwidth}{ \caption{\small{A summary of the weights and in/out of sample RMSEs for portfolios of 1 and 2 futures contracts which attempt to replicate a leveraged benchmark with $\beta=-3$. By comparison, the $-3$x LETF, DGLD has an out of sample RMSE of only 4.43718.}}\label{tab:Static Replication Table DGLD}}
    \end{small}
\end{table}

In contrast to the case of tracking the gold spot with futures (see Table \ref{tab:Static Replication Table}), the static portfolios do not replicate the leveraged benchmark well here. In Tables \ref{tab:Static Replication Table UGL} through \ref{tab:Static Replication Table DGLD}, the RMSE values are quite large for all the portfolios. The minimum RMSE for any portfolio of futures trying to replicate any leveraged benchmark is 22.64952 (achieved by a portfolio of 2-month and 6-month futures attempting to replicate a +2x investment in gold) and by comparison the maximum RMSE for any LETF trying to replicate its respective leveraged investment is 6.08133. (This is achieved by UGLD, which tracks a +3x investment in gold.) Unlike the unleveraged investment, the money market account is extensively used throughout the various portfolios. This is interesting but also logical. Indeed, in order to create leverage, the portfolio must either borrow if $\beta>0$ or invest in the money market account if $\beta<0$.

Furthermore, the optimal weights tend to lead to over/underleveraging. Since we are considering an investment in gold, the sum of the weights on the futures (which are proxy's for investment in gold) can be interpreted as the leverage on the portfolio. Since all the weights sum to 1, we can compute the approximate leverage as $1-w_0$. For the +2x and +3x investments, these values are all larger than 2 and 3, respectively. For the $-2$x and $-3$x investments, these values are all smaller (in absolute value) than $-2$ and $-3$, respectively. Granted, these are different instruments than the spot, but we have seen in Section \ref{futuresregression} they all move in parallel to the spot. Thus we see that the long portfolios tend to be over leveraged, while the short portfolios tend to be under leveraged. 

The optimization procedure has led to some rather uneven portfolio weights. For example, the optimal portfolio of 6-month and 12-month futures that attempts to replicate a +3x investment in spot gold requires the following transactions at inception: borrow \$6,140.11 from the money market account, short \$147,044.08 in 6-month futures and long \$154,184.19 in 12-month futures. In practice this would not be possible in the marketplace due to position limits that may be in place. 

\subsection{Dynamic Leverage Replication}\label{dynamicrepletfs}
To improve upon the replication in Section \ref{staticrepletfs}, we now consider a dynamic portfolio of one futures contract. Let $P_t$ be our portfolio value at time $t$. At every point in time, the portfolio invests $\beta$ times the value of the fund in the futures contract in order to achieve the required leverage. To fund this investment, the portfolio must borrow from the money market account. As a result, the value of our portfolio has the following dynamics:

\begin{equation}\label{portfolioevolution}
{dP}_t=\beta P_t \frac{dF_{t}}{F_{t}} - P_t (\beta-1) r_t dt.
\end{equation}
Here, $F_{t} $ is the value of a gold futures contract at time $t$ with a chosen maturity, e.g. 1 month, and $r_t$ is the risk free rate at time $t$.

\begin{remark}For our empirical analysis, we    need not  specify a parametric model for the futures price. Nevertheless, if one models the futures price by the stochastic differential equation
\[ \frac{dF_t}{F_t}= \mu_t dt + \sigma_t dW_t,\]with some stochastic drift $(\mu_t)_{t\ge 0}$ and volatility $(\sigma_t)_{t\ge 0}$, (many well-known models, including the  Heston, and exponential Ornstein-Uhlenbeck models, fit within  the above framework) then the log-price of the portfolio is given by
\begin{equation}
    \log P_t  =\log P_0 + \beta \log \frac{F_t}{F_0}  + \frac{\beta-\beta^2}{2}\Sigma_t+ (1-\beta)R_t ,
    \label{LETF Growth}
\end{equation}
where  $\Sigma_t= \int_0^t \sigma_s^2 ds$ is  the realized variance of $F$ accumulated up to time $t$ and $R_t=\int_0^t r_s ds$. Therefore,  under this general diffusion model, the log-return of the portfolio is proportional to the log-return of the futures by a factor of $\beta$, but also proportional to the variance by a factor of $\frac{\beta-\beta^2}{2}$. The latter factor is negative if $\beta \notin (0,1)$, which is true for every LETF considered here. \end{remark}

To implement the leveraged portfolio in \eqref{portfolioevolution}, we choose the front month futures contract. Recall from Section \ref{staticrepfutures} that the front month futures is the most effective in replicating the spot gold price. To calculate our portfolio value on each trading day, we discretize   \eqref{portfolioevolution} in time, using $\Delta t$  equal to 1 trading day and set $P_{0}$ equal to \$1000 as before.

To quantify our portfolio's replicating ability we will use the same root mean squared error that we used in Section \ref{staticrepletfs}:

\begin{equation}\label{rmsecalculationdynamicletf}
RMSE =  \sqrt{\frac{1}{n} \sum_{j=1}^n \left({P_j-L_j}\right)^2},
\end{equation}
where, $L_j$ is the value of a leveraged investment in gold and $P_j$ is the value of our portfolio, each at trading day $j$. For this dynamic portfolio, there is no sample from which we will need to draw our weights or train our model in any way. Therefore we can look at any conceivable time period and compare how the LETF $(L)$ or leveraged portfolio $(P)$ performs using the metric in \eqref{rmsecalculationdynamicletf}. 

For this tracking metric, we consider the period 1/3/2012 (first trading day of 2012) to 7/14/2014. The results are shown in Table \ref{tab:dynamicpfcumreturn}. The portfolio RMSEs range between 0.687\% and 3.291\%, which are smaller than LETF RMSEs which range between 1.87\% and 4.338\%. Overall, we see  that the portfolio RMSEs are lower than the LETF RMSEs. Indeed, we see that our dynamic portfolio is able to track the target leveraged index quite well according to the RMSE values for $\beta \in \{2,-2,3\}$. However the tracking is not as strong for $\beta=-3$. Nonetheless, the value is quite small and not that far off from the LETF RMSE. 

\begin{figure}[H] \centering
    \subfigure[Futures Portfolio]{\includegraphics[width=6in]{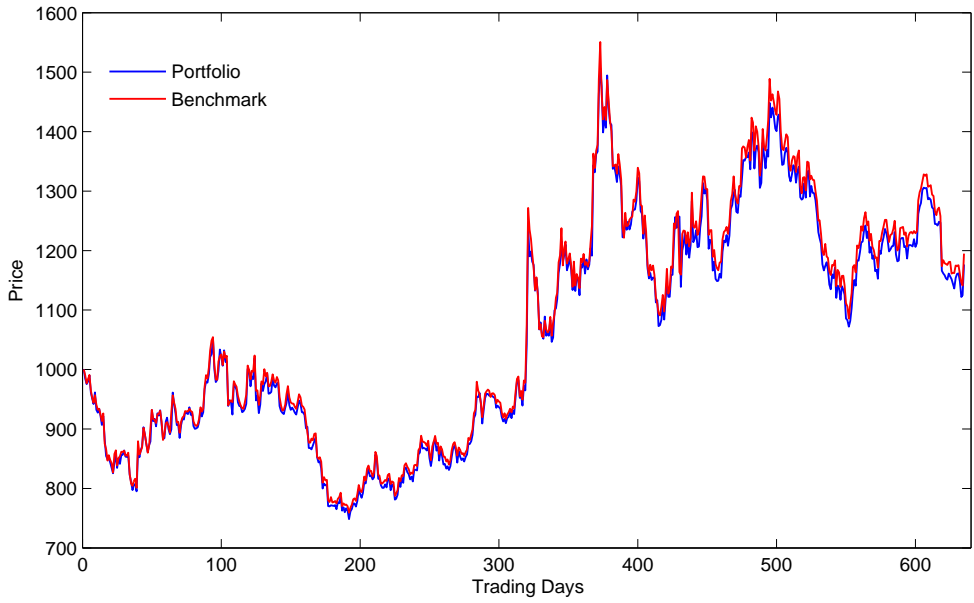}}
    \subfigure[GLL]{\includegraphics[width=6in]{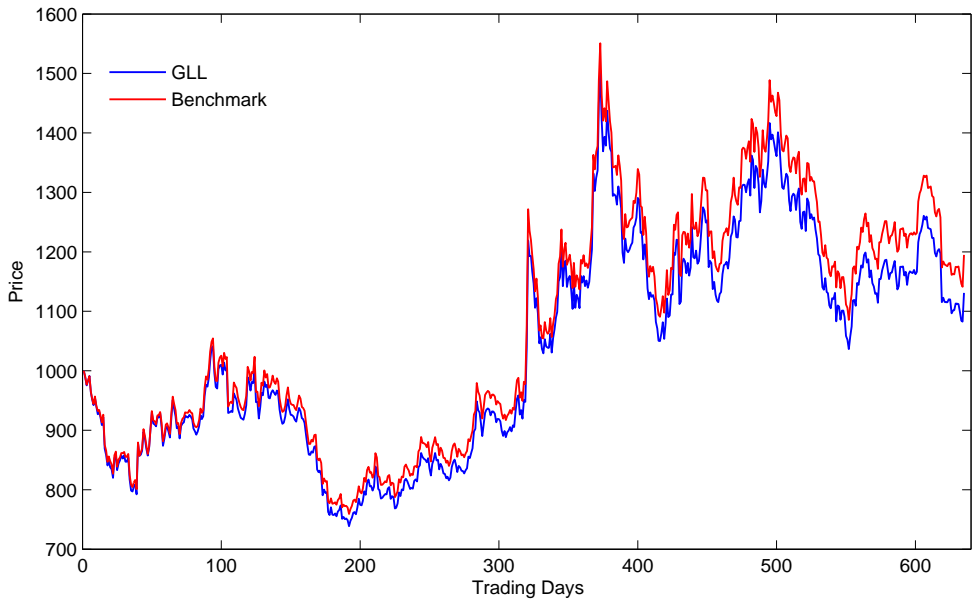}}
    \parbox{0.85\textwidth}{\caption{\small{Time evolution of our dynamic portfolio of front month futures and cash (top) compared to the $-2$x benchmark and GLL (bottom) compared to the $-2$x benchmark. Time period displayed is 1/3/2012 to 7/14/2014.}}\label{fig:dynamicrepfigure}}
\end{figure}

In Figure \ref{fig:dynamicrepfigure}, we see the time evolution for both the dynamic portfolio and GLL compared to the $-2$x benchmark. It is visible that the LETF tends to underperform the benchmark and the difference worsens over time. On the other hand, the portfolio tends to stay close to the benchmark over the entire period. Though not reported here,   we  observe similar patterns for other gold LETFs.

In Table \ref{tab:dynamicpfcumreturn}, we also give the annual returns for each asset for the years 2011, 2012 and 2013. For UGLD and DGLD we do not have data for the full year of 2011 (its issue date was 10/17/2011) so we do not have annual returns for these LETFs in 2011. The dynamic portfolio returns range between -69.22\% and 107.54\% while the LETF returns range between -69.90\% and 106.16\%. Comparing each year and leverage ratio pair, we find that, except for $\beta=-3$ in 2012, our portfolio outperforms the LETF in each year. Thus, we have shown that in general, a dynamic portfolio consisting of just one futures contract can not only more closely track the target leveraged index, but it also outperforms the respective LETF. \\

\vspace{30pt}

\begin{table}[H]\centering\begin{small}
    \setlength{\extrarowheight}{2pt}
    \begin{tabular}{c c r r r r}
    	\hline
    	& & & \multicolumn{3}{c}{Annual Returns (\%)} \\ 
        \hline
        $\beta$ & \text{Asset} & $RMSE$ & {2011} & {2012} & {2013} \\
        \hline
        \hline
        \text{+2x} & {UGL} 	& 30.33 & 12.90 & 2.81 & -52.31\\
        			& {Portfolio}      	& 6.87   & 15.23 & 6.29 & -51.83 \\
	\hline
        \text{$-2$x} & {GLL} 	& 40.12   & -29.43 & -16.40 & 67.82\\
        			& {Portfolio}      	& 15.87     & -27.06 & -14.89 & 70.57\\
	\hline
        \text{+3x} & {UGLD} 	& 43.38   & {-} & 0.41 & -69.90\\
        			& {Portfolio}      	& 12.55   & {-} & 5.29 & -69.22\\
	\hline
        \text{$-3$x} & {DGLD} 	& 18.70   & {-} & -23.57 & 106.16\\
        			& {Portfolio}      	& 32.91   & {-} & -24.51 & 107.54\\
       \hline    \end{tabular}
  \parbox{0.8\textwidth}{ \caption{\small{A summary of the annual returns (over the periods: 1/3/2011-12/31/2011, 1/3/2012-12/31/2012 and 1/2/2013-12/31/2013) and RMSE for each LETF and a dynamic portfolio of front month futures and cash. RMSE values are calculated over the period 1/3/2012-7/14/2014.}}\label{tab:dynamicpfcumreturn}}
    \end{small}
\end{table}

%

\section{Concluding Remarks}\label{conclusion}

In this paper, we  have studied the price relationship among gold spot, futures, ETFs, and leveraged ETFs.  There are significant price co-movements among the spot, futures and ETF (GLD). We show that   static portfolios consisting 1 or 2 futures with different maturities can effectively  replicate the spot gold price. As for   leveraged  gold ETFs,   their average returns tend to be lower than the corresponding multiple ($\beta =3, 2, -2, -3$)  of the  spot's returns, and the under-performance worsens  over a  longer holding period.  In order to track the leveraged benchmark, we construct a dynamic leveraged  portfolio using the 1-month futures. We  demonstrate the portfolio tracks the leveraged benchmark better than the corresponding LETFs, and has better returns over multiple years. 
 
As the ETF market continued to grow in quantity and complexity in various asset classes, it is important for both investors and regulators to understand the price dependency among unleveraged and leveraged ETFs, as well as their underlying assets and related derivatives. We have explored futures contracts on gold as a replicating instrument for gold (L)ETFs. It would therefore be important to continue the study of the gold futures market  and its impact on the dynamics of LETFs. 

As for other future research, there are also options written on gold (L)ETFs. This calls for  consistent pricing of LETF options across   leverage ratios  (see   \cite{leungsircarLETF}). Also, one can analyze the market impact (see e.g. \cite{DA}) of the (L)ETF issuer on the particular derivative market used to replicate the (L)ETF. As we have noted before, (L)ETFs have large market capitalization and thus, an adjustment of the issuer's portfolio could lead to large movements in the derivatives market used to hedge the (L)ETF. The risks associated with (L)ETFs, futures and other derivatives need to be studied in depth in order to help advise and motivate policies issued by the Fed (see e.g. \cite{CMAA} for a study of the risk of VIX futures).

Finally, studies that can  shed  light on  the risks of LETFs (see e.g. \cite{LeungMarco2012}), and their  impact on the broader financial market will be very useful. In the recent study, 
\cite{SK} examine the performance of LETFs  during the financial crisis.   We have seen LETFs on gold have failed to track the desired leveraged target. There  are a number of reasons that may contribute to the tracking errors including, but not limited to, illiquidity and   risks of  the hedging instruments, management fees and financing costs  (see e.g. \cite{Charaput}). Studies that illuminate the significance of each of these effects  would be useful for   investors as well as regulators.

\begin{small}
    \bibliographystyle{apa}
    \bibliography{mybib}
\end{small}

\end{document}